\documentclass[twoside,slac_one]{revtex4}
\usepackage{graphicx}
\usepackage{fancyhdr}
\usepackage{amsmath} % American Mathematics Society standards
\usepackage{bm}% bold math
\usepackage{amsxtra}
\usepackage{amssymb}
\usepackage{amsthm}
\usepackage{latexsym}
\usepackage{lscape}
\usepackage{subfig}

\begin{document}

\title{SciBath: A Novel Tracking Detector for Measuring Neutral 
Particles Underground}


\author{R. Cooper, L. Garrison, H.-O. Meyer, T. Mikev, L. Rebenitsch,
R. Tayloe}
\affiliation{Department of Physics, Indiana University, Bloomington, IN 47405, USA}

\begin{abstract}
The SciBath-768 detector is a prototype neutral particle detector
offering high-precision reconstruction of neutrino and neutron events.
It consists of a three dimensional grid of 768 wavelength-shifting fibers
immersed in 82 liters of liquid scintillator.
Initially conceived as a charged particle detector for neutrino
studies, it is also sensitive to fast neutrons (1-100~MeV).  Simulation results
show 30\% efficiency and 30\% energy resolution for 1-10~MeV tagged
neutron events. 
The apparatus has been commissioned and will be deployed in Fall 2011
to measure neutrinos and neutrons 100 meters underground in the
Fermilab MINOS near-detector area.
\end{abstract}

\maketitle

\thispagestyle{fancy}


%%%%%%%%%%%%%%%%%%%%%%%%%%%%%%%%%%
\section{Motivation}

%%%%%%%%%%%%%%%%%%%%%%%%%%%%%%%%%%
\subsection{Neutrino Interactions}
The original motivation for the SciBath detector was 
to reconstruct 1~Gev neutrino neutral-current elastic (NCel) scattering events, allowing a direct 
measurement of the strange-quark contribution to the spin
of the nucleon~\cite{FINeSSE}. The experimental signature for this interaction is a single low energy proton
of typically 100~MeV kinetic energy, which has a range of approximately 10~cm in liquid scintillator.
Accurate reconstruction of these 10~cm proton tracks is difficult in a 10~ton scintillator detector, 
which the small neutrino cross section (typically $\sigma \approx 10^{-38}$~cm$^2$) requires.

\begin{figure}[ht]
\centering
\includegraphics[width=80mm]{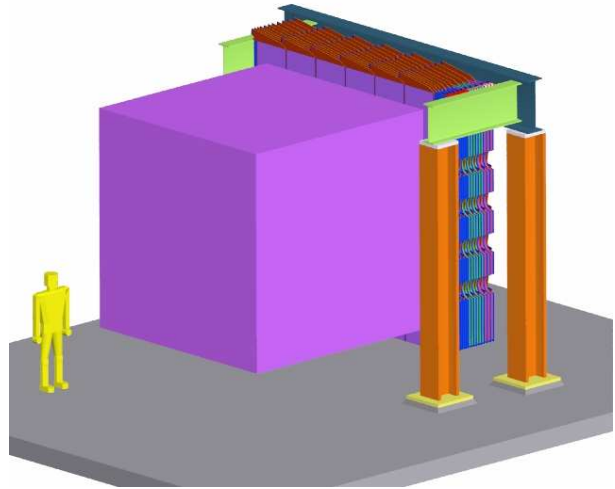}
\caption{The FINeSSE detector as modeled in a GEANT3~\cite{GEANT3} simulation.  The front cubic region contains a 
19k-fiber SciBath detector. A muon range stack is located behind.} 
\label{Figure1}
\end{figure}

The detector described here, SciBath-768, is a prototype of the vertex detector 
proposed for the Fermilab Intense Neutrino Scattering Scintillator 
Experiment (FINeSSE)~\cite{FINeSSE}, 
which was was designed to measure these 10~cm proton tracks using the SciBath technology. 
The full-sized detector proposed consisted of 19,200 wavelength-shifting fibers immersed in liquid 
scintillator of volume $(2.5$~m$)^3$.
This design would increase the performance/price ratio for neutrino detectors
and would allow the economic construction of a large, high-resolution
device.

%%%%%%%%%%%%%%%%%%%%%%%%%%%%%%%%%%
\subsection{Neutron Detection}
Ambient neutron fluxes are an important background for some experiments, especially
low-rate underground experiments searching for dark matter or neutrinoless double-beta
decay.  However, these fluxes are difficult to predict.
The SciBath-768 detector could aid these experiments by directly measuring
fast neutron (1--100~MeV) fluxes.
In general, fluxes of cosmic-ray muon induced neutrons have only been
measured at a few sites~\cite{Mei}.
SciBath-768 could be used to experimentally map out these fluxes as a
function of depth and at a particular experiment location.

%%%%%%%%%%%%%%%%%%%%%%%%%%%%%%%%%%
\section{Principle of Operation}
The SciBath design starts with a light-tight container filled with liquid scintillator.
Wavelength-shifting (WLS) fibers are arranged with three sets of parallel, mutually 
orthogonal fibers arranged in a grid.
Scintillation light from charged particles moving through this container is
captured by the embedded wavelength-shifting fibers and is guided to
multi-anode photomultiplier tubes (MAPMTs).
The amount of light captured depends on the distance between the
particle track and the capturing fiber in a well-understood fashion.
Thus, by measuring the amount of light captured by each fiber the parameters 
of the particle track can be reconstructed.
The symmetry of this fiber arrangement allows reconstruction of particle
tracks at arbitrary angles with good efficiency.
The energy deposited by a charged particle can be interpreted from the total
number of photons recorded by all fibers.

%\begin{figure}[ht]
%\centering
%\includegraphics[width=80mm]{figure2.eps}
%\caption{A schematic view of the SciBath fiber arrangement showing the three orthogonal
%fiber orientations.} \label{fig:fibgrid}
%\end{figure}

%%%%%%%%%%%%%%%%%%%%%%%%%%%%%%%%%%
\section{Previous Work}

%%%%%%%%%%%%%%%%%%%%%%%%%%%%%%%%%%
\subsection{FINeSSE Simulations}
\begin{figure}[ht]
\centering
\subfloat{\includegraphics[width=0.45\textwidth]{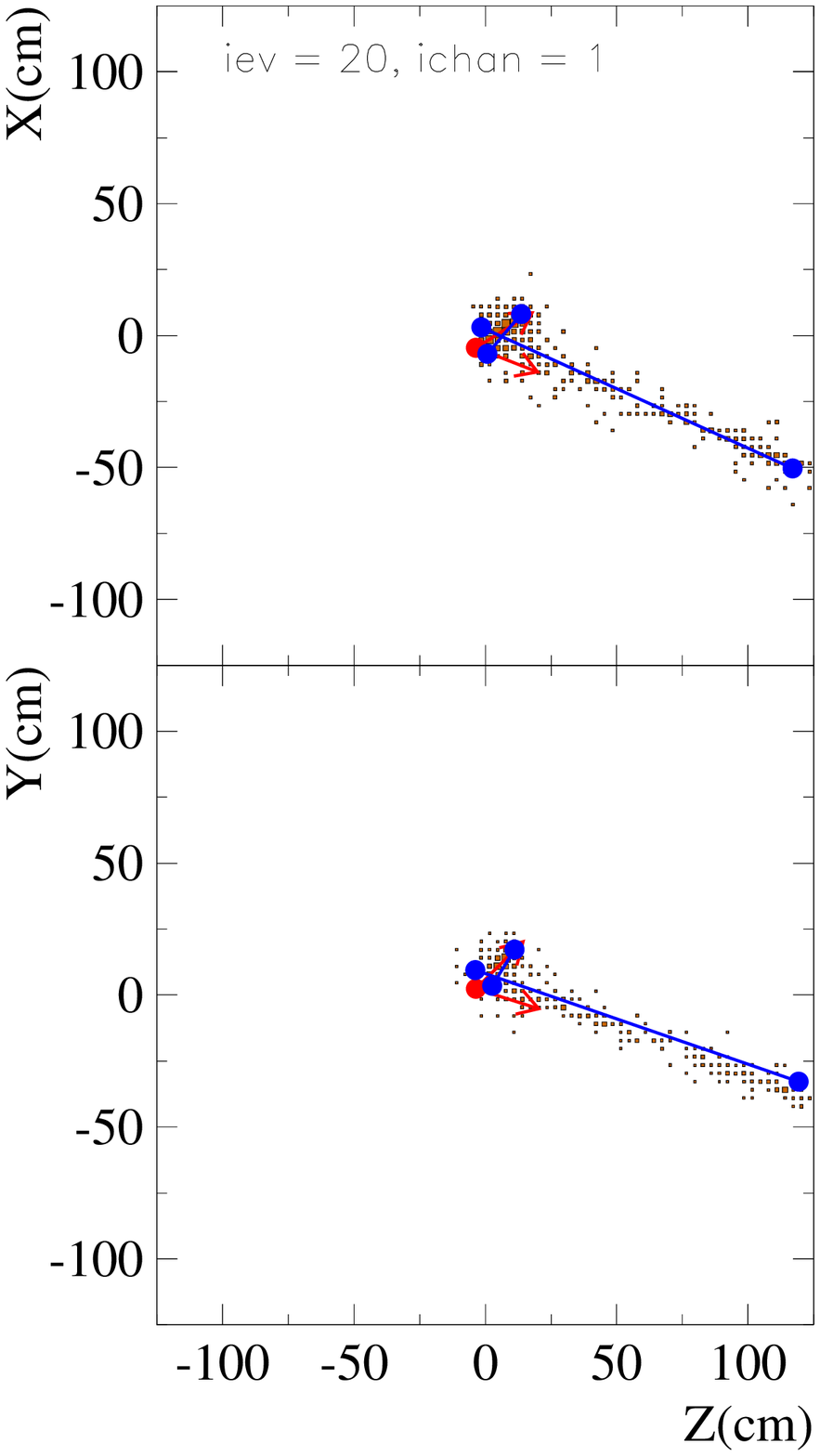}}
\subfloat{\includegraphics[width=0.45\textwidth]{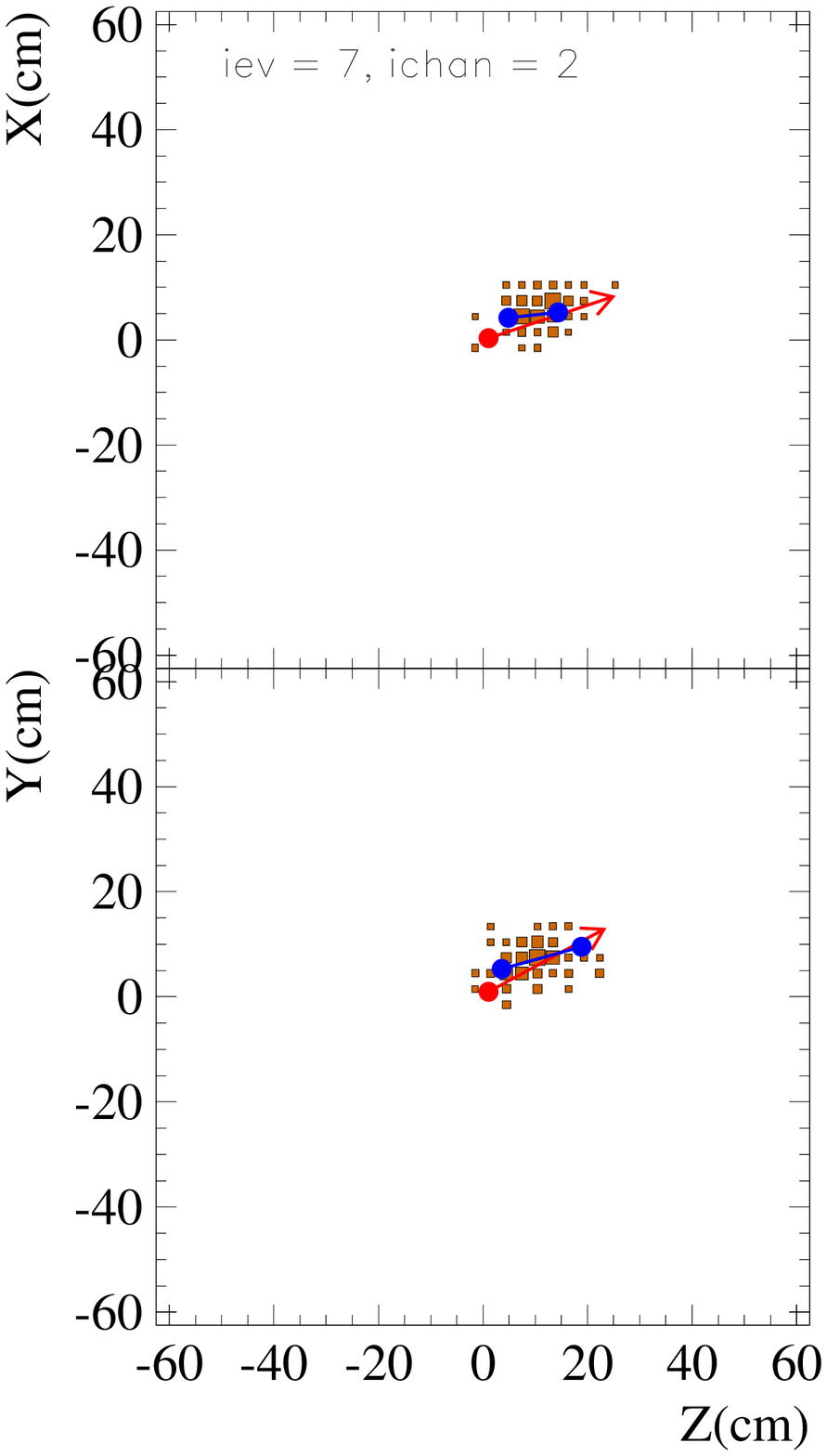}}
\caption{Event displays from a GEANT3 simulation of neutrino scattering events in the 
FINeSSE SciBath detector. The top plots show the XZ (top) projection and the bottom plots 
YZ (side) projections. A CCQE event with a muon and proton in the final state is shown 
on the left and a NCel event with a single final-state proton is shown on the right.  The boxes
indicate individual fiber hits, the lines terminating with dots are the reconstructed
tracks, and the lines with arrows show the true tracks and vertex.}
\label{fig:FINevdisp}
\end{figure}

A GEANT3~\cite{GEANT3} simulation of the FINeSSE detector showed excellent 
reconstruction of neutrino events with energy and angular resolution of approximately
10~MeV and 100~mrad, respectively, for the physics events of interest. 
An event display from this simulation for a charged current
quasi-elastic (CCQE) and a  neutral current
elastic (NCel) event is shown in Figure~\ref{fig:FINevdisp}.
In the NCel event, the single recoil proton is visible and well-reconstructed. 
The cross section from the CCQE event can be used along with the cross 
section for NCel scattering to generate a more accurate measurement of the
strange quark contribution to the spin of the proton than could be obtained
using the NCel scattering cross section alone.
The CCQE events are easier to identify because the resulting muon creates a
longer, more obvious track.

%%%%%%%%%%%%%%%%%%%%%%%%%%%%%%%%%%
\subsection{SciBath-30}
\begin{figure}[ht]
\centering
\includegraphics[width=0.7\textwidth]{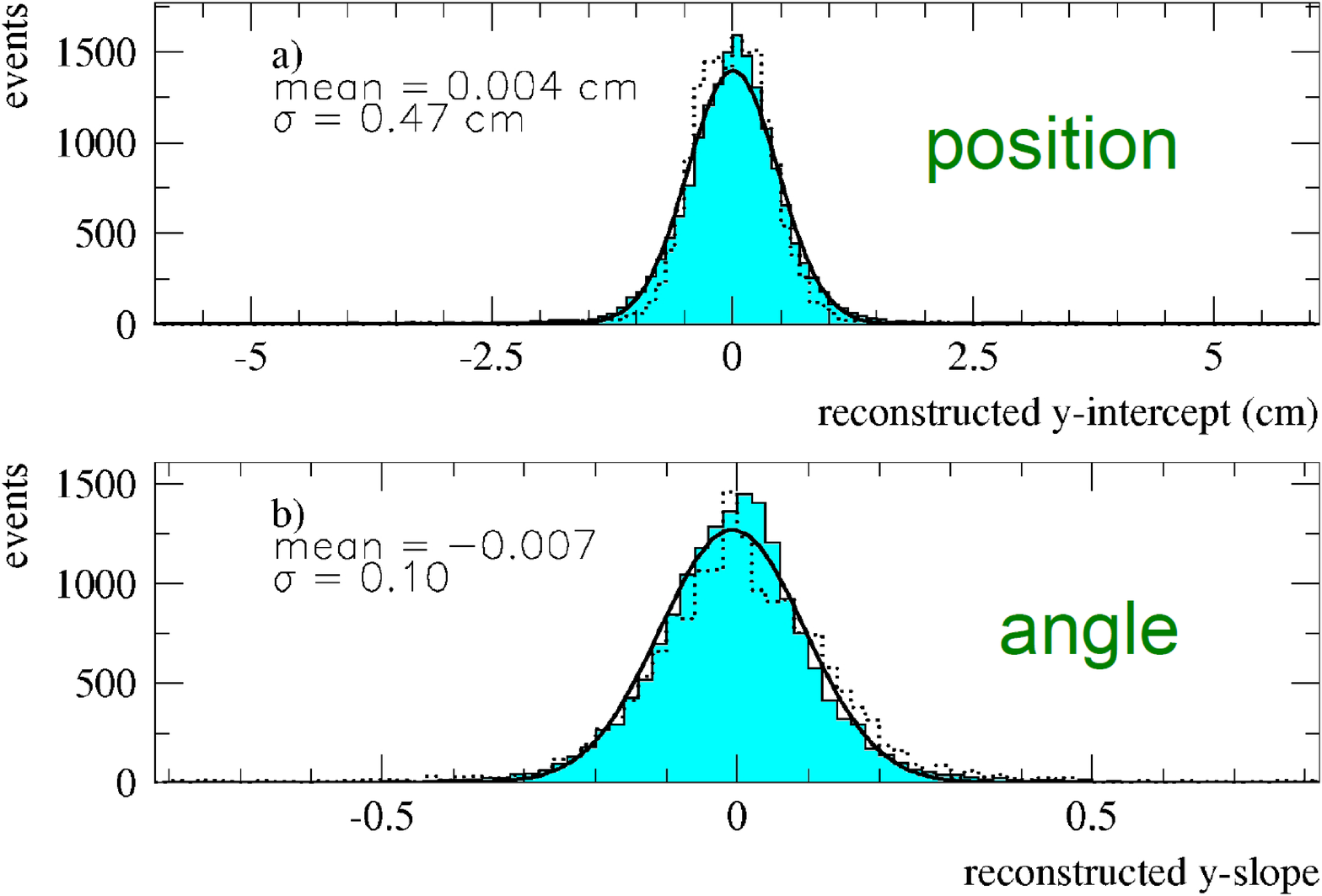}
\caption{Position (a) and angular resolution measured in the SciBath-30 device.}
\label{fig:sb30res}
\end{figure}

SciBath-30 was a ``proof-of-principle'' device that was used to demonstrate
the viability of the SciBath technology~\cite{scibathNIM}. It contained 30 
parallel wavelength-shifting fibers immersed in liquid scintillator.
This device was tested in the Indiana University Cyclotron Facility 200~MeV
proton beam.  These tests allowed for the tuning of the optimal WLS fiber/liquid 
scintillator combination and a measurement of the position and angular resolution for
200~MeV protons. Figure~\ref{fig:sb30res} shows the resulting  5~mm position and 
6$\,^{\circ}$ angular resolution, quite adequate for the FINeSSE experiment.

%\begin{figure}[ht]
%\centering
%\subfloat[Design]{\includegraphics[width=0.4\textwidth]{figure4a.eps}}
%\subfloat[Position and Angular Resolution]{\includegraphics[width=0.4\textwidth]{figure4b.eps}}
%\caption{SciBath-30}
%\label{Figure4}
%\end{figure}

%%%%%%%%%%%%%%%%%%%%%%%%%%%%%%%%%%
\section{SciBath-768}

%%%%%%%%%%%%%%%%%%%%%%%%%%%%%%%%%%
\subsection{Detector Description}
\begin{figure}[ht]
\centering
\includegraphics[width=0.5\textwidth]{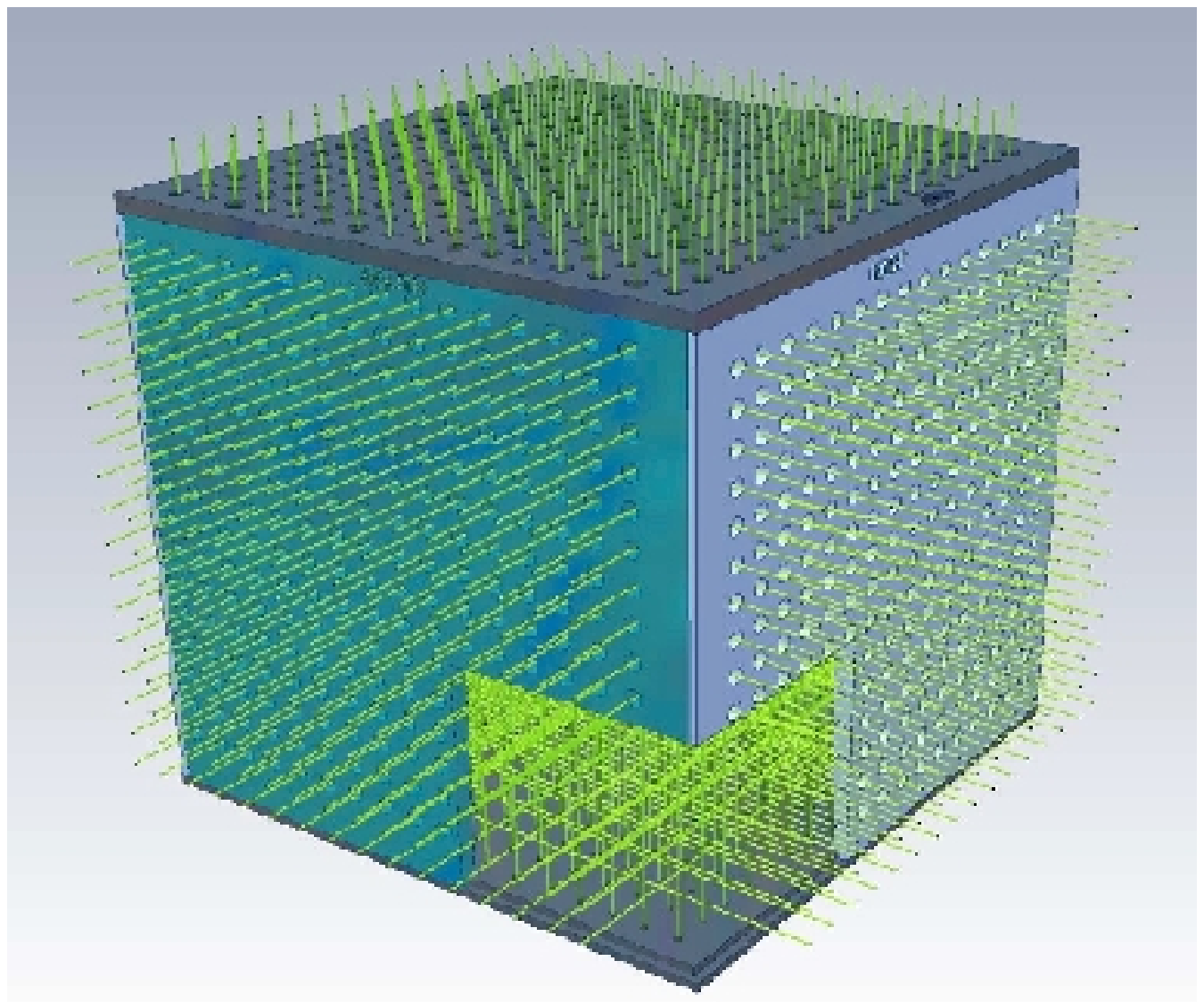}
\caption{A schematic view with cutaway of the SciBath-768 cube and WLS fibers.} 
\label{fig:sb768cube}
\end{figure}

SciBath-768 is a prototype of the proposed FINeSSE experiment
and consists of a (45~cm)$^3$ cube containing 82 liters of liquid scintillator 
and 768 WLS fibers as shown in  Fig.~\ref{fig:sb768cube}. 
The scintillator is a custom mix of mineral oil with  11\% pseudocumene, and 
1.5~g/l diphenyloxazole (PPO) with an emission wavelength in the 350--400~nm range.

\begin{figure}[ht]
\centering
\subfloat[Overview showing cube, LED calibration system, and VME crates]
{\includegraphics[width=0.45\textwidth]{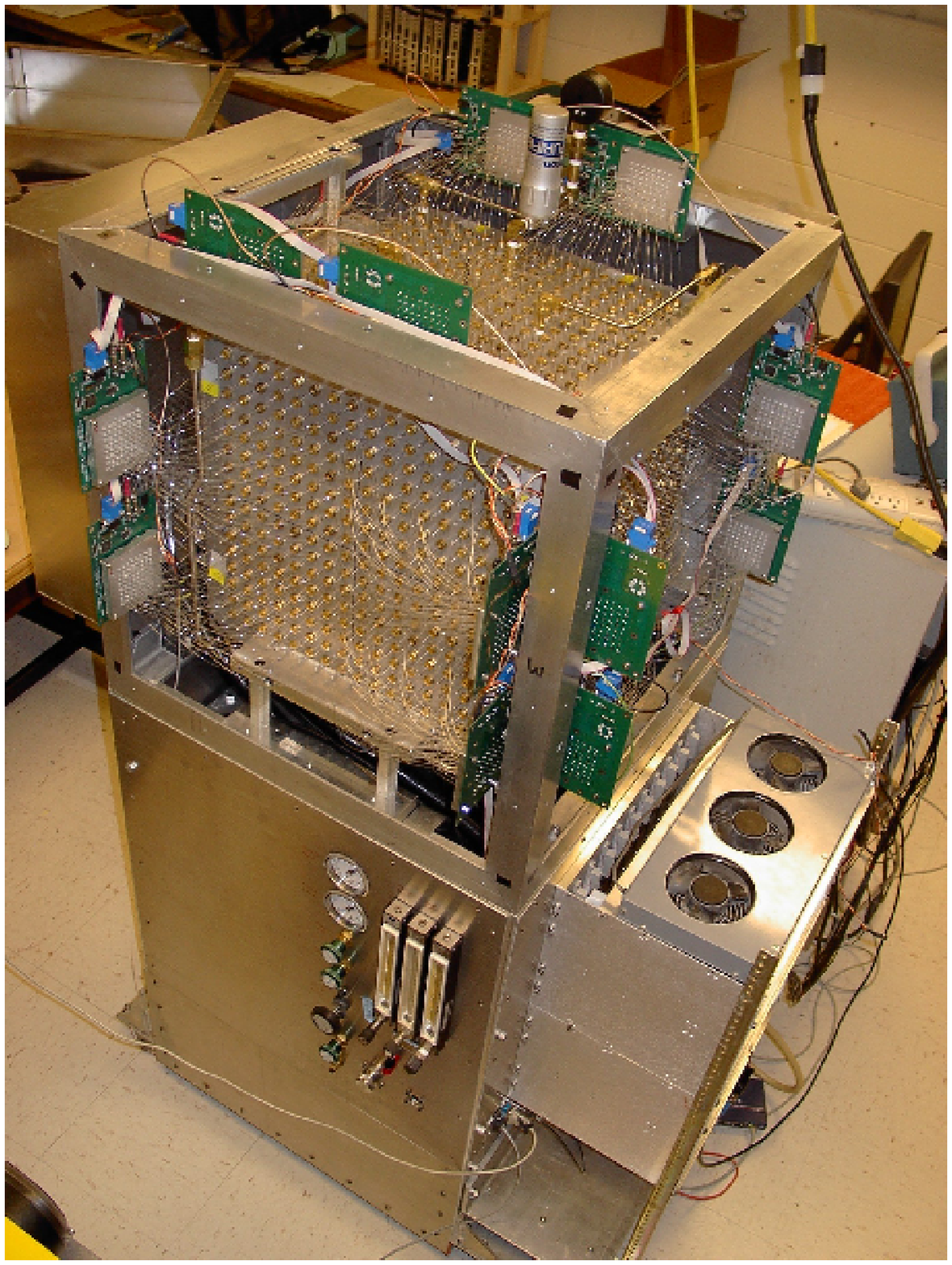}}
\subfloat[LED calibration system]
{\includegraphics[width=0.45\textwidth]{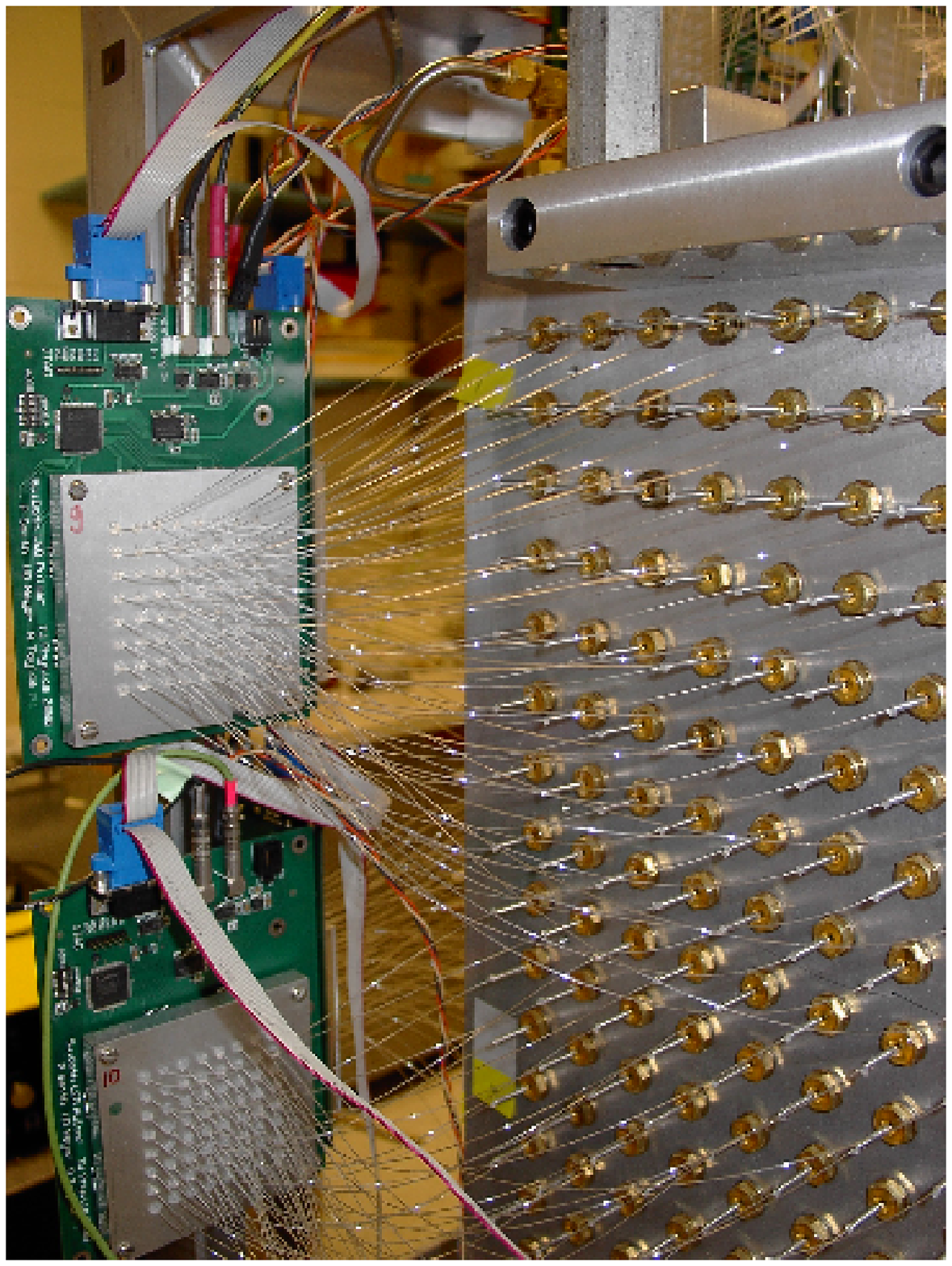}}
\caption{SciBath-768 photos}
\label{fig:sb768photos}
\end{figure}

The WLS fibers~\cite{WLSfibs} are arranged in three 16x16 grids that are
mutually orthogonal with a 2.5~cm fiber spacing. They are 
1.5~mm in diameter and shift light from ultraviolet (320--370~nm) to blue (410--480~nm).
Both ends of each fiber protrude outside the cube where one end couples to
1.5~mm diameter clear plastic optical fibers that routes the wavelength-shifted light
to MAPMTs. The other exposed end of each wavelength-shifting fiber is coupled to a
custom-built pulsed LED calibration system with one LED per fiber (see Fig.~\ref{fig:sb768photos}).

\begin{figure}[ht]
\centering
\subfloat[Integrated Readout Module with PMT]{\includegraphics[width=0.45\textwidth]{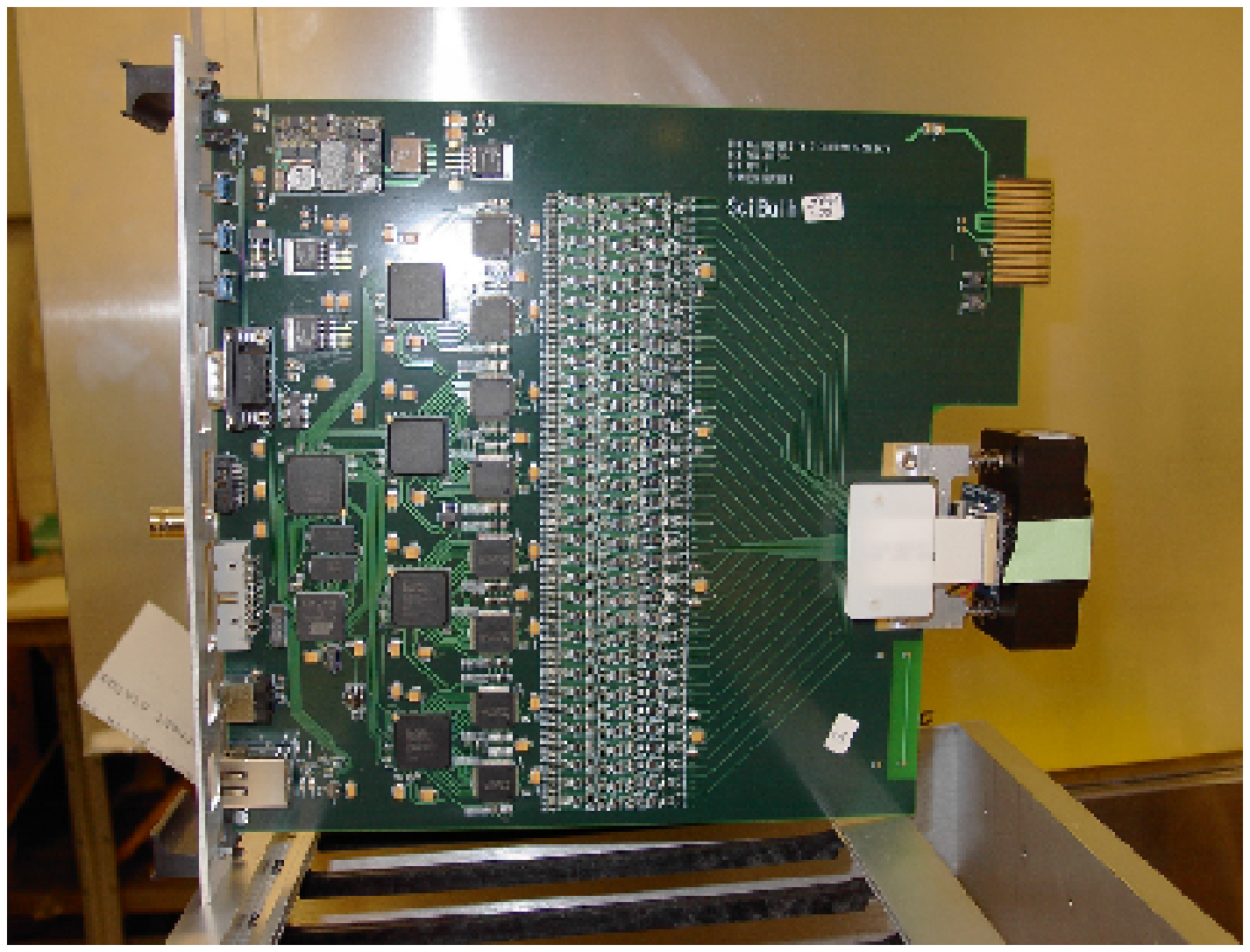}}
\subfloat[PMT waveform]{\includegraphics[width=0.5\textwidth]{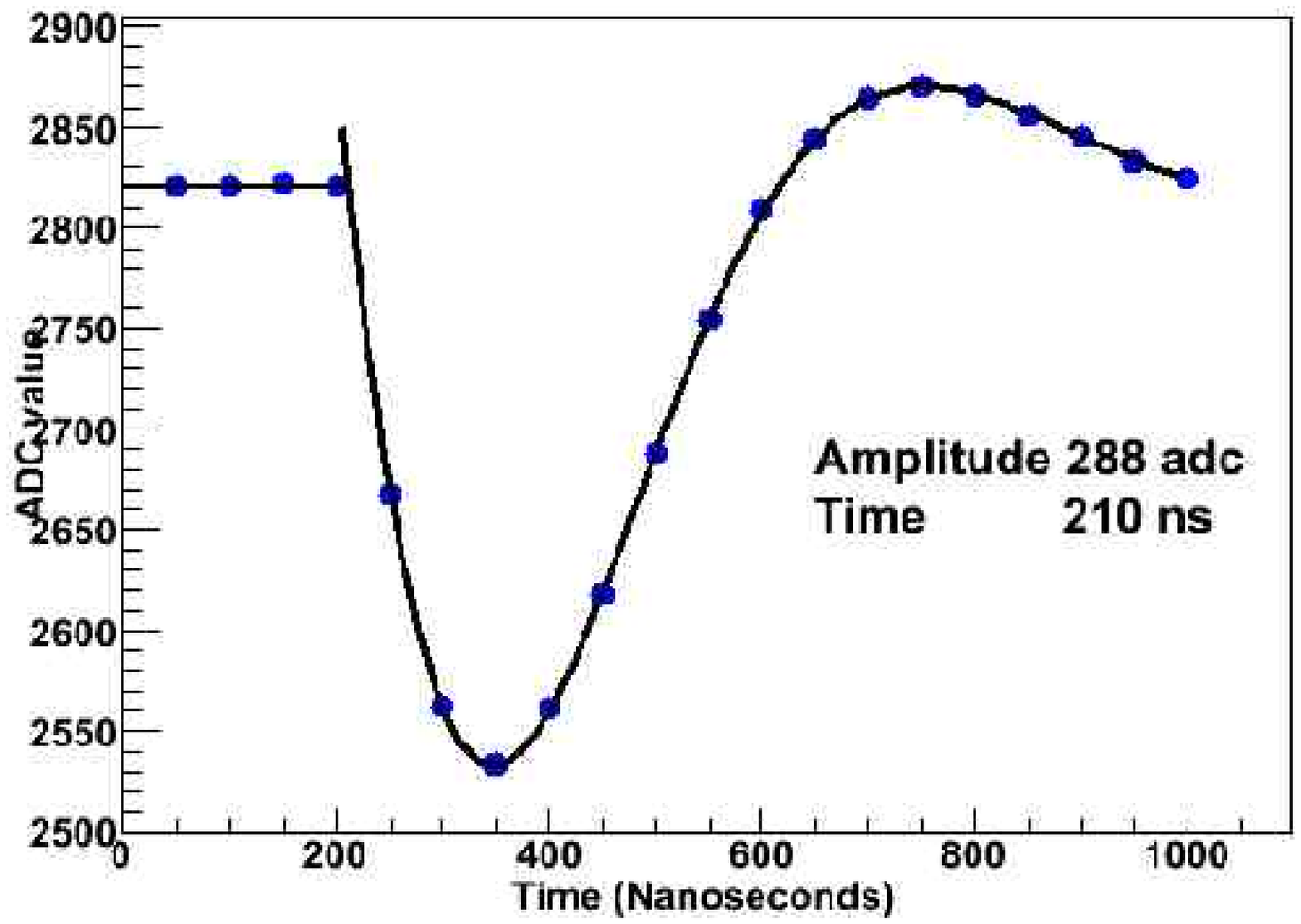}}
\caption{SciBath-768 readout}
\label{fig:IRM}
\end{figure}

The readout electronics consist of 12 custom-built ``Integrated
Readout Modules'' (IRMs), each with an integrated Hamamatsu (R7600) 64-anode
MAPMT~\cite{Hamamatsu}. These IRMs are located on the detector in two VME crate ``shells''.
Each IRM (Fig.~\ref{fig:IRM}a) utilizes flash analog-to-digital converters (ADCs) 
sampling at 20~MHz, five field programmable gate arrays (FPGAs), and an ARM9 microcontroller.
The final cost of these readout electronics is on the order of \$70/channel 
(including the MAPMT).

The MAPMT signals are first shaped by a ``ringing oscillator'' circuit as a front-end
in the IRM to allow a time and charge measurement with one ADC channel. 
A typical MAPMT waveform as input to the flash ADC is shown in Figure~\ref{fig:IRM}b.
The FPGAs provide zero suppression and buffering before data is shipped off
via ethernet to a data acquisition computer for analysis.

%%%%%%%%%%%%%%%%%%%%%%%%%%%%%%%%%%
\subsection{Current Status}
The SciBath-768 detector and data acquisition (DAQ) system are currently assembled 
and operating. 
The first iteration of analysis and simulation software are also finished and
in use. We are currently commissioning and calibrating with cosmic muon events
and the LED system. 

\begin{figure}[ht]
\centering
\subfloat[Simulated Muon Track]{\includegraphics[width=3.4in]{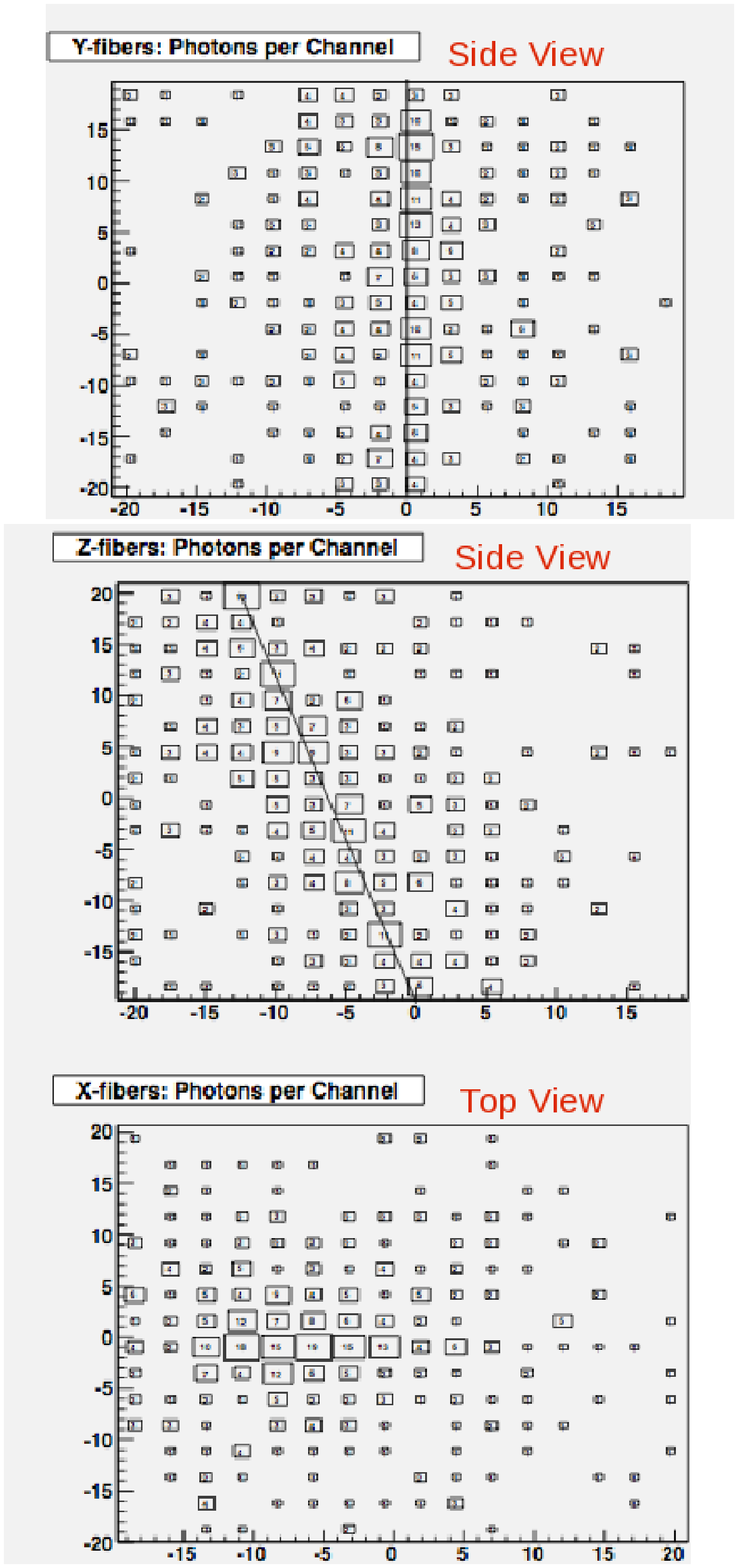}}
\subfloat[True Muon Track]{\includegraphics[width=3in]{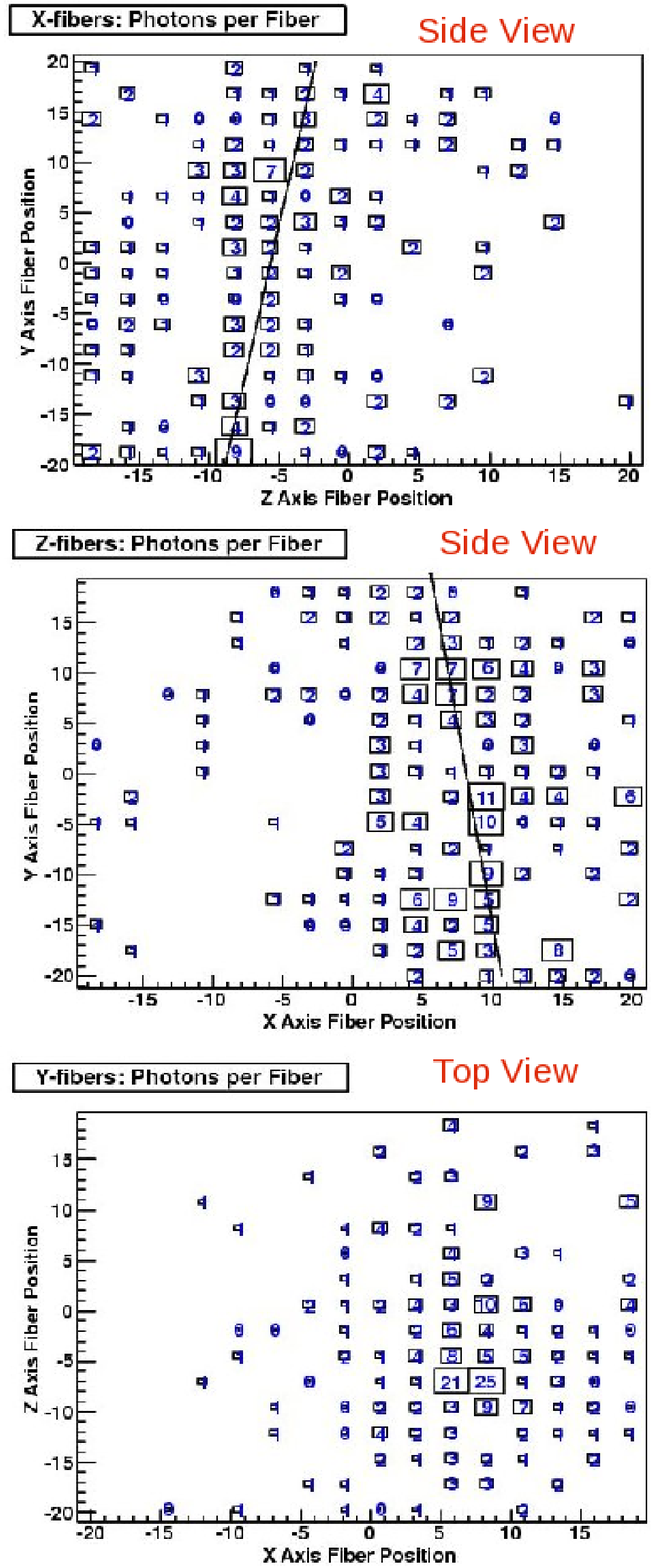}}
\caption{SciBath-768 event display}
\label{fig:sb768evdis}
\end{figure}

Figure~\ref{fig:sb768evdis} shows both a simulated and a detected (true) cosmic muon event.
Here each WLS fiber (and PMT channel) records a ``pixel'' on the event display, 
where the box (pixel) size is proportional to the number
of photons seen by that fiber. Note that in this figure only a preliminary 
calibration has been applied.

Ongoing work includes further calibration with cosmic muon and LED data, upgrading
DAQ software, and developing particle identification algorithms.
We are also preparing for a three month run this fall at Fermilab
in the MINOS near-detector area.

%%%%%%%%%%%%%%%%%%%%%%%%%%%%%%%%%%
\subsection{Simulations}
The detector is currently simulated with a GEANT4~\cite{GEANT4} code which
is used to predict the detector response to various incident particles. 
Cosmic muons typically result in a through-going track in the SciBath-768 detector.
These tracks  are easy to identify and provide a sample to use for simulation
and reconstruction tuning.  

Using this simulation, a track reconstruction method was developed using
the method of least squares for perpendicular offsets~\cite{MathWorld} and the MINUIT~\cite{MINUIT} 
minimization routine in ROOT~\cite{ROOT}. This method demonstrated 3~mm position resolution 
and 5$\,^{\circ}$ angular resolution for through-going muon tracks.

A neutron in our scintillator will predominantly scatter elastically on protons 
generating short recoil proton tracks losing energy until it thermalizes and is 
captured via the n(p,d)$\gamma$ reaction.
This neutron capture generates a signature 2.2~MeV photon in a characteristic
time of 186~$\mu$s.  This characteristic signal of a prompt recoil protons followed
by neutron capture may be used to tag stopping neutron events. 

\begin{figure}[ht]
\centering
\subfloat[Distribution of the number of detected photons in SciBath-768 
from a simulated 2.2~MeV neutron capture signal]
{\includegraphics[width=0.55\textwidth]{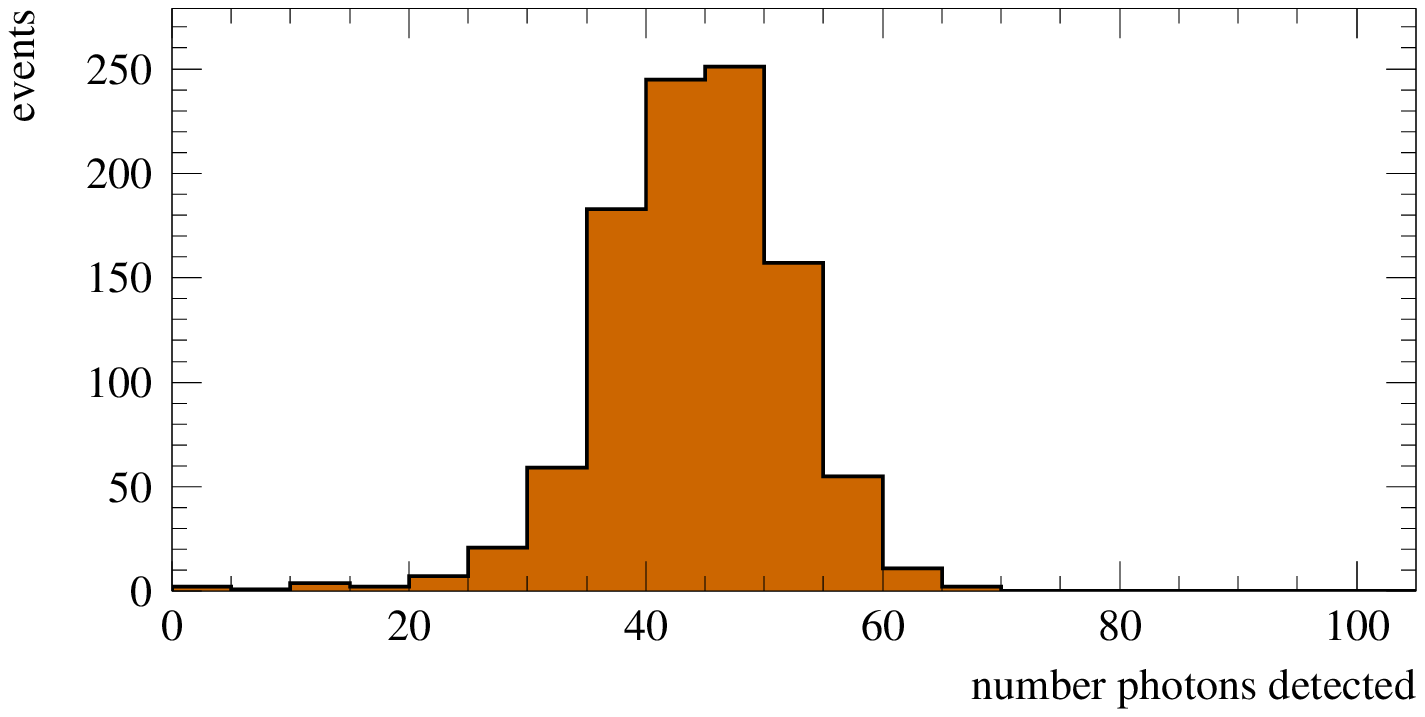}}
\subfloat[Simulated tagged neutron detection efficiency and energy resolution
for 1--100~MeV (kinetic) neutrons.]
{\includegraphics[width=0.4\textwidth]{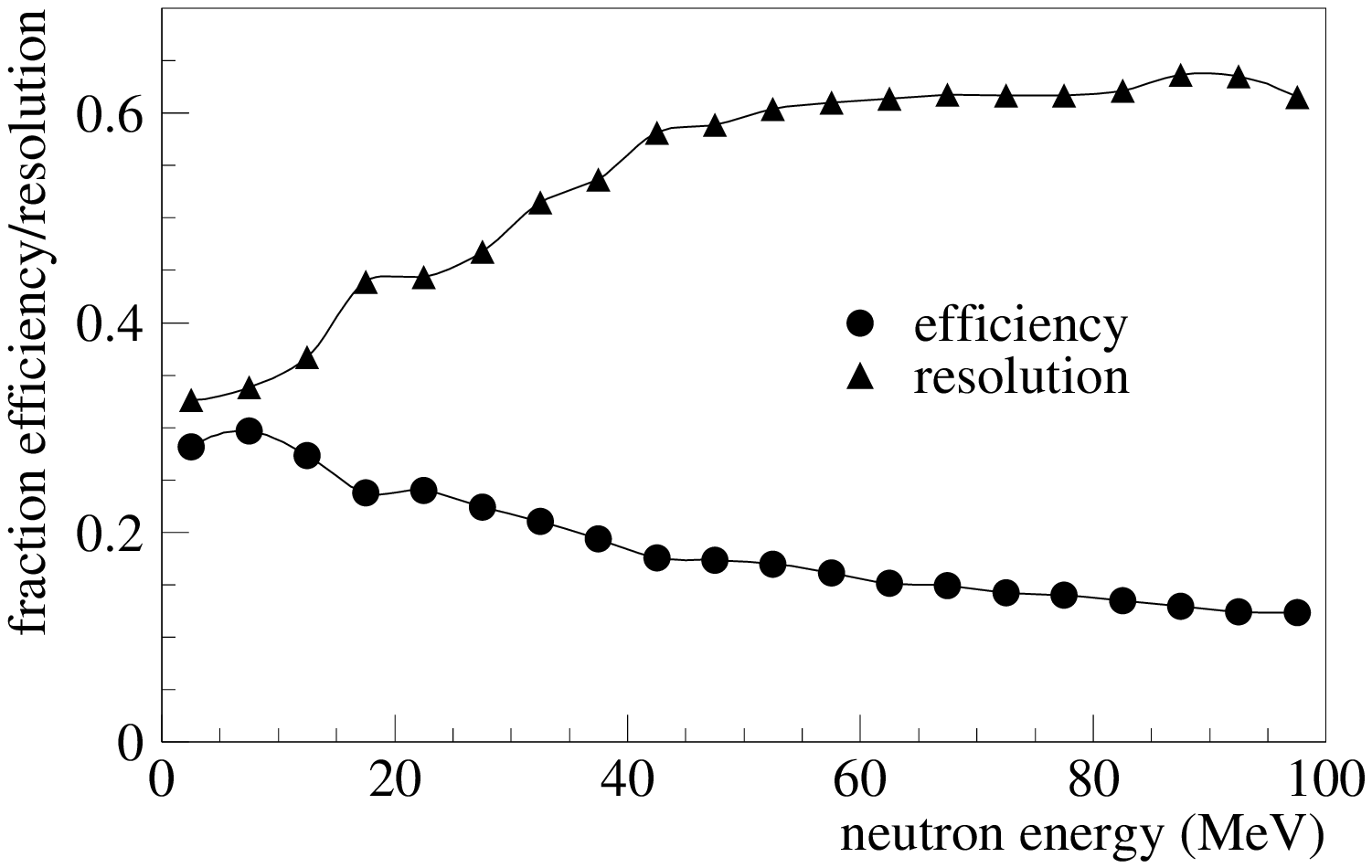}}
\caption{SciBath-768 tagged neutron simulation results.}
\label{fig:sb768sim}
\end{figure}

A simulation of the response of SciBath to a 2.2~MeV photon from a neutron
capture event leads us to expect roughly 45 photons detected, as shown in
Figure~\ref{fig:sb768sim}(a).   Further simulation has shown that SciBath-768 
will be able to use this signature to tag 1--100~MeV neutrons with efficiency
and energy resolution as shown in Figure~\ref{fig:sb768sim}(b).  In the 1--10~MeV
energy range a 30\% efficiency and 30\% energy resolution are predicted.
This evolves to approximately 10\% efficiency and 60\% energy at 100~MeV
as it is more probable to lose some fraction of the neutron energy due
to escaping particles.

%%%%%%%%%%%%%%%%%%%%%%%%%%%%%%%%%%
\subsection{Future Plans}
\begin{figure}[ht]
\centering
\includegraphics[width=\textwidth]{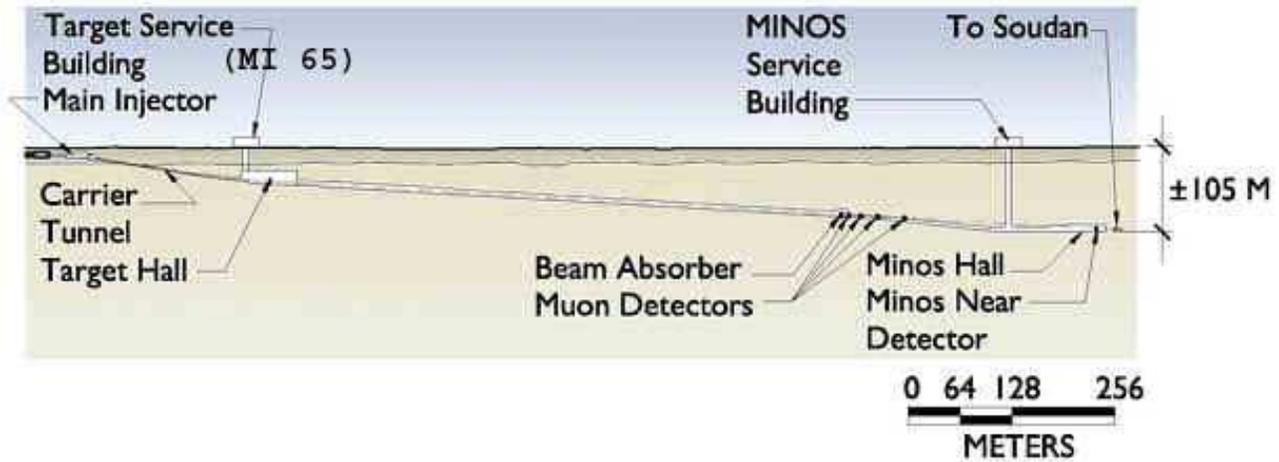}
\caption{An elevation view of the NuMI tunnel and MINOS near-detector area at Fermilab.
The SciBath-768 detector will be located in the MINOS Hall just upstream of the
MINOS near-detector.}
\label{fig:NuMI}
\end{figure}

In the Fall of 2011, the SciBath-768 detector will be moved to and run 100~m underground 
in  the MINOS near detector hall at Fermilab (Fig.~\ref{fig:NuMI}).  
The goals for this run are threefold.
First, the NUMI beam will allow for a demonstration of neutrino event reconstruction.
In the three months that we plan to run at FNAL (two month livetime) we
expect to see between 100 and 10000 neutrino events depending on the neutrino
beam configuration (Table~\ref{tab:numievs}).

\begin{table}[ht]
\begin{center}
\begin{tabular}{|l|c|c|}
\hline \textbf{Beam Configuration} & \textbf{$\nu$ CC Inclusive} & 
\textbf{$\nu$ CCQE} \\
\hline Neutrino, low energy & 550 & 100 \\
\hline Neutrino, medium energy & 12000 & 1400 \\
\hline Antineutrino: low energy & 200 & 30 \\
\hline Antineutrino: medium energy & 4000 & 1300 \\
\hline
\end{tabular}
\caption{Expected $\nu$ / $\bar{\nu}$ events in SciBath-768 in the
MINOS near-detector hall in 2 months for all possible beam configurations.  ``Neutrino''
and ``Antineutrino'' refer to the meson selection in the neutrino production target,
producing a predominantly neutrino or antineutrino beam. The terms ``low energy'' and 
``medium energy'' refer to the energy selection of the neutrino beam, with ``low'' providing
a peak energy of 3~GeV and ``medium'' 6~GeV.} 
\label{tab:numievs}
\end{center}
\end{table}

The second goal of the run is to measure the cosmic-ray-induced fast (1--100~MeV)
neutron flux. From estimates based on FLUKA simulations~\cite{Mei}, expected events
rates are 20 cosmic neutron events per day.
The third goal is to measure the neutrino-beam induced neutron flux in this
location.

After a successful demonstration of the capabilities of SciBath-768 at
Fermilab, we will next evaluate options to run in an underground lab at
a greater depth to measure the fast neutron flux background relevant for 
future underground experiments.

\end{document}